\documentclass[a4paper,11pt,twocolumn]{article}

\usepackage{times}
\usepackage{epsfig}
\usepackage[american]{babel}
\usepackage{graphicx} 

\begin{document}

\title{ The production of proton and lepton fluxes in near Earth orbit}

\author{P. Zuccon$^1$,
B. Bertucci$^1$,
B. Alpat$^1$,
R. Battiston$^1$,
G. Battistoni$^2$,
W.J. Burger$^1$, \\
G. Esposito$^1$,
A. Ferrari$^{2,3}$,
E. Fiandrini$^1$,
G. Lamanna$^{1,}$\footnote{Now at CERN - Switzerland} \ ,
P.R. Sala$^{2,3}$}

\date{January 17, 2002}



\onecolumn

\maketitle

\begin{center}

$^1$ Universit\`a and Sezione INFN of Perugia, Italy

$^2$ Sezione INFN of Milano, Italy

$^3$ CERN Switzerland

\end{center}

\begin{abstract}
Substantial fluxes of protons and leptons with energies 
below the geomagnetic cutoff have been measured by
the AMS experiment at altitudes of 370-390 Km, in the 
latitude interval $\pm$51.7$^o$. The production mechanisms of 
the observed trapped fluxes are investigated in detail 
by means of the FLUKA  Monte Carlo simulation code. 
All known processes
involved in the interaction of the cosmic protons 
with the atmosphere (detailed descriptions of the magnetic field 
and atmospheric density, as well as the electromagnetic and 
nuclear interaction processes) are included in the simulation. 
The results are presented and compared with the experimental data,
indicating good agreement with the observed fluxes.
The impact of secondary proton flux on particle production in atmosphere
is briefly discussed.
\end{abstract}

\section{Introduction}
Cosmic rays approaching the Earth interact with the atmosphere resulting in
a substantial flux of se\-condary particles. The knowledge of composition, 
intensity and energy spectra of these particles is of considerable interest,
e.g. for the evaluation of background radiation for satellites and 
the estimate of the atmospheric neutrino production for neutrino 
oscillation experiments \cite{batti}. 

The AMS measurements in near earth orbit \cite{AMSpra,AMSele} have allowed, 
for the first time, to gather accurate information on the intensity, energy 
spectra and geographical origin of charged particle fluxes at energies below
the geomagnetic cutoff over a wide range of latitudes 
and at almost all longitudes. 
The under cutoff component of proton fluxes at equatorial latitudes 
has revealed an unexpected intensity of up to $50\%$ of the primary proton 
flux, a positron to electron flux ratio has been found in the undercutoff 
component which largely exceeds the cosmic one, differences in residence 
times and geographical origins have been reconstructed for positively and 
negatively charged particles. 

A robust interpretation of these and many other characteristics
of the undercutoff fluxes in terms of secondary particles produced 
in atmosphere requires an accurate description of both the interaction processes at their origin and of the 
geomagnetic field effects.
Recently, different interpretations of the AMS measurements have been
proposed \cite{derome,Plyaskine} based on Monte Carlo simulations using 
different approaches on both the generation technique and the 
interaction model.

In this work, we report results from a fully 3D Monte Carlo simulation
based on FLUKA 2000 \cite{fluka} for the description of cosmic ray 
interactions with the atmosphere. The key features of our analysis
are an efficient generation technique for the incoming proton flux 
and a true microscopic, theory driven treatment of the interaction 
processes opposite to empirical parametrization of accelerator data.
As a first attempt the contribution of $He$ and the heavier nuclei,
representing ($\approx 9\%$) \cite{bartol} of the all nuclei cosmic 
flux,  is neglected. 

In the following section we give a detailed description of the  basic
ingredients of this simulation, the generation technique and the interaction
model. In section 3 we present the results on both protons and leptons 
and the  comparison with AMS measurements. In section 4 we propose our
conclusions.

\section{The model}
An isotropic flux of protons is uniformly generated  on a geocentric
spherical surface with a radius of 1.07 Earth radii ($\sim 500~\!Km$ a.s.l.)
in the kinetic energy range $0.1\!-\!170~GeV$.

We took the functional form suggested in \cite{hkkm} to describe the proton
energy spectrum, the spectral index and the solar modulation parameter  are 
extracted from a fit to the AMS data \cite{AMSprb}.

The magnetic field in the Earth's proximity includes  two
components:
the Earth's magnetic field, calculated using a 10 harmonics IGRF \cite{IGRF}
implementation, and the external magnetic field, calculated using the
Tsyganenko Model\footnote{The external magnetic field is calculated
only for distances  greater than 2 Earth's radii ($E_{R}$) from the Earth's
center . Its contribution to the total magnetic field
is $<\!1\%$ at smaller distances and therefore can be safely neglected.}
\cite{tsyga}. To account for the geomagnetic effects, 
for each primary proton we back-trace an antiproton of the same energy 
until one of the following conditions is satisfied: 
\begin{enumerate}
\item the particle reaches the distance of $10~\!E_{R}$ from the
Earth's center. 
\item the particle touches again the production sphere.
\item neither 1 or 2 is satisfied before a time limit is reached.
\end{enumerate}

If condition 1 is satisfied  the particle is on an allowed trajectory,
while if condition 2 is satisfied the particle is on a forbidden one.
Condition 3 arises for only a small fraction of the events
$O(10^{-6})$.

Particles on allowed trajectories are propagated forward and can reach the
Earth's atmosphere.
The atmosphere around the Earth is simulated up to $120~
Km$ a.s.l.   using 60 concentric layers of homogeneous density and chemical
composition. Data on density and chemical composition are taken from the
standard MSIS model \cite{msis}. 
The Earth  is modeled as a solid sphere which absorbs each particle 
reaching its surface.

\subsection{ The generation technique}

The  ideal  approach in the generation of the primary cosmic
rays spectra would be to 
start with an isotropic distribution of particles at a great distance
(typically $10~\!E_{R}$)
from the Earth where the geomagnetic field introduces negligible distortions 
on the interstellar flux. 
However, this
computational method is intrinsically inefficient since  most of
the particles are  generated 
with trajectories which will not reach the Earth environment. 
Kinematic cuts can be applied in order to improve the selection efficiency
at generation, however they tend to introduce a bias particles with low
rigidity.

A good alternative to this approach is the 
 backtracing
method \cite{derome},\cite{Lipari} adopted in the present analysis as
outlined in 
the previous section. In the following, we will shortly discuss the validity
of the technique and report the results of  a comparison of the
two methods. 
We recall  that this method was applied for the first time 
in ref. \cite{hkkm} for the generation of atmospheric neutrino fluxes.

Let us consider first the effects of the geomagnetic field on an 
incoming flux of charged particles in the absence of a solid Earth. 
For the discussion, we start with an isotropic flux of monoenergetic\footnote
{The realistic case of an energy spectrum can be treated just as
a superposition of monenergetic cases}
protons at large 
distance, i.e. at infinity, from the origin of a geocentrical reference 
system. In this scenario, a negligible fraction of particles, with
very particular initial kinematic parameters, will follow complicated paths
and 
remains confined at a given distance from the origin (semi-bounded 
trajectories); for all practical purposes this sample can be ignored. 
Most of the particles will follow unbounded trajectories, 
reaching again infinity after being deflected by the magnetic field.

\begin{figure}[htb]
\center
\includegraphics[width=8.5cm]{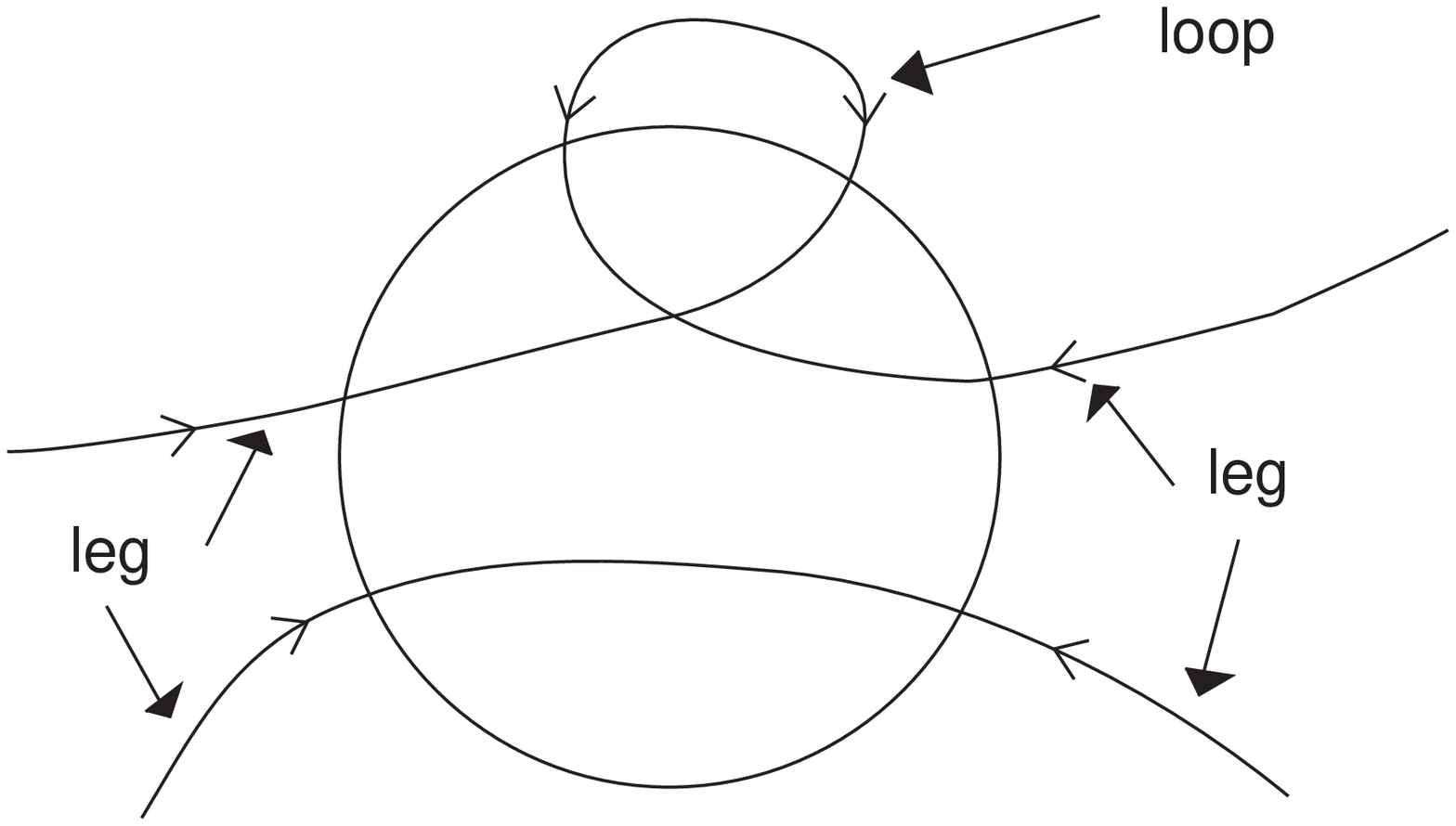}
\caption{Trajectories types crossing a spherical surface around the Earth}
\label{fig_med1}
\end{figure}

Unbounded trajectories  cross a spherical surface centered in the field
source only an even number of times, as shown in Fig.\ref{fig_med1}: we  call
{\it legs} the trajectory parts connecting the spherical surface to infinity
and {\it loops} the parts of the trajectory starting and ending inside the
spherical surface. 

Since each trajectory can be followed in both directions and no source or sink
of particles is contained within the surface, the incoming and outgoing fluxes
are the same. However, the presence of the magnetic field breaks the
isotropy of the flux ``near'' the field source, so for a
given location there is a flux dependence  due to the direction.

Applying the Liouville Theorem, under  the hypothesis of
isotropy at infinity, it is straightforward to prove \cite{vallarta} that 
 the proton flux in a random point is the same as at infinity
along a set of directions (allowed directions), and zero along all the others
(forbidden directions).

The pattern of the allowed and forbidden directions depends on both
the rigidity and the location and is  known as the geomagnetic cutoff.

\begin{figure}[htb]
\center
\includegraphics[width=8.5cm]{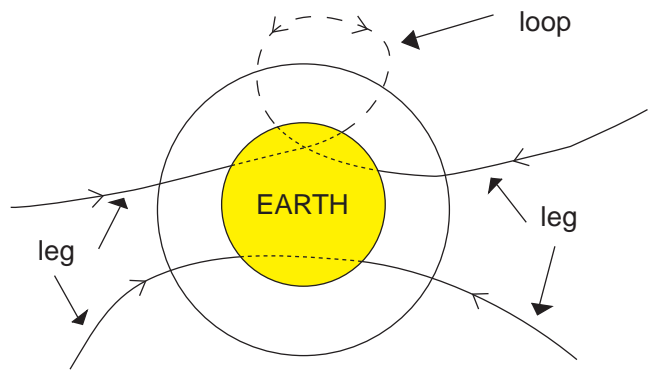}
 \caption{Trajectories in the presence of a solid Earth}
\label{fig_med2}
\end{figure}

With the introduction of a solid Earth, all the trajectories that are crossing
the Earth are broken in two or more pieces (Fig.\ref{fig_med2}): the {\it legs}
become  one-way trajectories and the {\it loops} disappear.

The presence of the Earth modifies the flux which exits from 
the surrounding spherical surface, since  particles are absorbed
by the Earth,
while it has only a minimal effect on the incoming flux which is modified only
by the absence of certain {\it loops}. 
To generate the flux of
particles reaching the Earth's atmosphere, it is sufficient to follow the
particles along the allowed trajectories corresponding to the {\it legs},
taking care to avoid
double or multiple counting.

\begin{figure}[!hbt]
\center
\includegraphics[width=6.5cm]{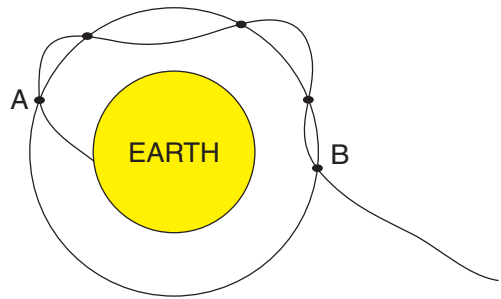}
 \caption{ An example of multiple counting along  a trajectory, this type
of trajectory has to be considered only at point B.}
\label{fig_med3}
\end{figure}

To respect this prescription we  reject all trajectories that
are back-traced to the production sphere, this allow us to correctly consider
the cases like the one shown in Fig.\ref{fig_med3}.

We point out  that an important difference with respect to the application
in the neutrino flux calculation of \cite{hkkm} is
that for the former,  the generation sphere coincided with the Earth's
surface, and therefore the forbidden trajectories included those which touched
again the Earth (plus those who remained trapped for a  long time). In that
case there are no problems of double counting.

To check the validity of our technique we made a test comparing  the results
of the inefficient generation technique at 10 Earth's radii distance
from the Earth's center with the backtracing technique described in this paper.

\begin{figure}[!hbt]
\center
 \includegraphics[width=\textwidth]{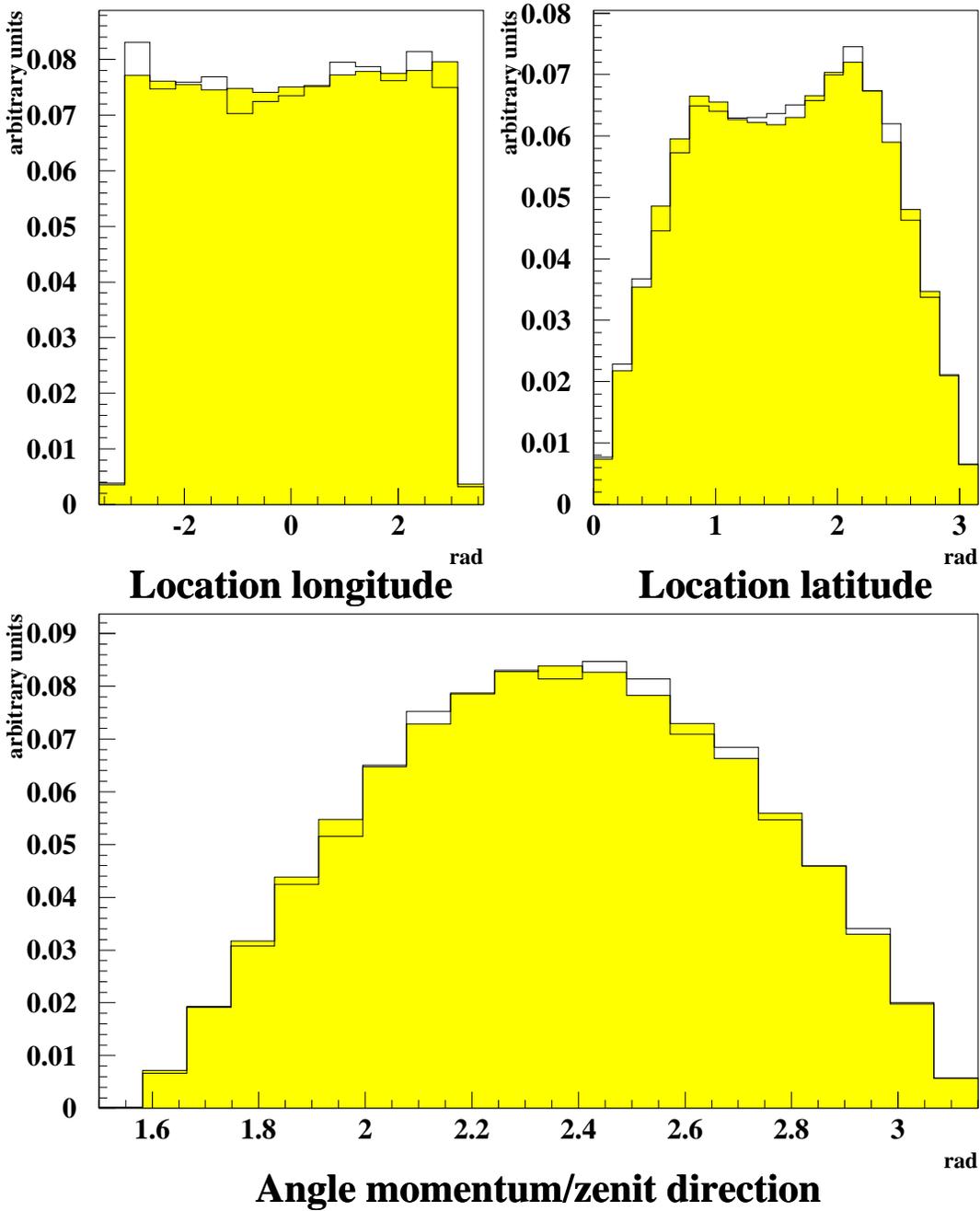}
 \caption{ Latitude and longitude of impact points and angle between
momentum and zenith directions for particles generated at a distance of 10
Earth's radii (solid line) and particles generated at 1.07 Earth's radii
(shaded histogram). }
 \label{fig_med4}
\end{figure}

Fig. \ref{fig_med4} shows this comparison for several cha-racteristic
distributions, the agreement between the two methods is good.

\subsection{ The interaction model}
We use the software package FLUKA 2000 \cite{fluka} to transport  the
particles and describe their interactions with Earth's atmosphere.
The setup of this simulation is derived from the one used in \cite{batti}.
This package contains a tridimensional description of both electro-magnetic
and hadronic interactions.
This code is benchmarked
against a wide set of data and is already used in many applications,
ranging from low energy nuclear physics to high energy accelerator and
cosmic ray physics. For this reason we have preferred this model with 
respect to the use of ``ad hoc'' parametrizations of particle production
in the energy range of our interest.

In FLUKA hadronic interactions are treated in a theory-driven approach, and
the models and their implementations are  guided and checked using experimental
data. 
Hadron-nucleon interaction models are based on resonance production and decay
below an energy of few $GeV$ and on the Dual Parton Model above. 
The extension from hadron to  hadron-nucleus interactions is done in the
framework of a generalized intra-nuclear cascade approach including the
Gribov-Glauber multi-collision mechanism  for higher energies followed by
equilibrium processes: evaporation, fission, Fermi break-up and $\gamma$
de-excitation.  The parameters of the models embedded in the FLUKA
package are fixed only by comparing expectations with data from accelerator
experiments.

\begin{figure}
\center
\includegraphics[width=0.6\textwidth]{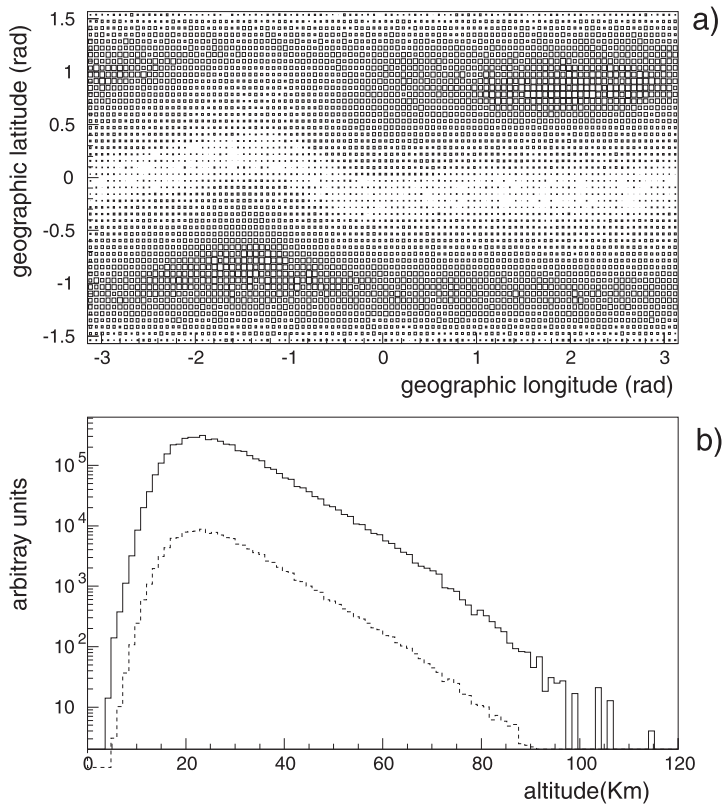}
\caption{ a) distribution of primary protons interaction points in
geographical coordinates, b)
altitude profile of primary protons interaction points, solid line $E_{k}\!<\!
30~GeV$, dashed line $E_{k}\!>\!30~GeV$ }
\label{fig1}
\end{figure}

In fig \ref{fig1} a) we show the map of the primary proton 
interaction points in geographical coordinates. The distribution reflects
the influence of the geomagnetic cutoff. 
Fig \ref{fig1} b) shows the interactions altitude profile,
the solid histogram is for $E_{k}\!<\!30~GeV$ while the dashed
one is for $E_{k}\!>\!30~GeV$. The mean interaction altitude
depends weakly on the energy. 

The cosmic proton impinging in the atmosphere are doing elastic scattering
in the  24\% of the events and inelastic interactions in the remaining
76\%, in tab.\ref{tab2}
we show some characteristic of the inelastic interactions as simulated by
FLUKA 2000.

\begin{table}[ptb]
\begin{center}
\begin{tabular}{|l||c|c||c|c||c|c||c|c|}
\hline
& \multicolumn{2}{c}{$5 GeV$} \vline \vline &
\multicolumn{2}{c}{$10 GeV$}\vline  \vline &
\multicolumn{2}{c}{$20 GeV$} \vline \vline &
\multicolumn{2}{c}{$30 GeV$} \vline \vline \\
\hline
Particle& Mult.& E frac.& Mult.& E frac.& Mult.& E frac.& Mult.& E frac.\\
\hline
\hline
$p$      &1.983  &0.409  &2.676  &0.337  &2.744  &0.307 &2.770 &0.294 \\
\hline
$\pi^{+}$ &0.711  &0.131  &1.292  &0.149  &1.970  &0.159 &2.381 &0.164\\
\hline
$\pi^{-}$ &0.389  &0.068  &0.975  &0.098  &1.641  &0.116 &2.047 &0.122\\
\hline
$\pi^{0}$ &0.638  &0.114  &1.601  &0.169  &2.378  &0.175 &2.840 &0.177 \\
\hline
\end{tabular}
\end{center}
\caption{Energy fraction and multiplicity of secondary particles for
the proton interactions with atmospheric nuclei in FLUKA 2000. 
Four  typical energies of primary protons are considered.}
\label{tab2}
\end{table}

\section{Comparison with the AMS data}

To compare with the AMS data, we define a detection boundary corresponding to
a spherical surface at the AMS orbit altitude ($400\,\mathrm{Km}$  a.s.l). We record 
each  particle that crosses the detection boundary within the AMS
field-of-view, defined as a cone with a 32$\,\mathrm{^o}$ aperture  with respect to
the local zenith or nadir directions. 

To obtain the absolute normalization, we take into account 
the field-of-view, the corresponding AMS acceptance, and an
Equivalent Time Exposure (E.T.E.) corresponding to the number of
the generated primary protons.

Our results are based on a sample of $\sim 6\cdot 10^{6}$ primary protons
 generated in the kinetic energy range of $0.1-170~GeV$, which corresponds 
to \mbox{$\sim 4~10^{-12}s$~
(E.T.E).} 

\subsection{Protons}

\begin{figure*}[htb]
\vskip -1cm
\center
 \includegraphics[width=\textwidth]{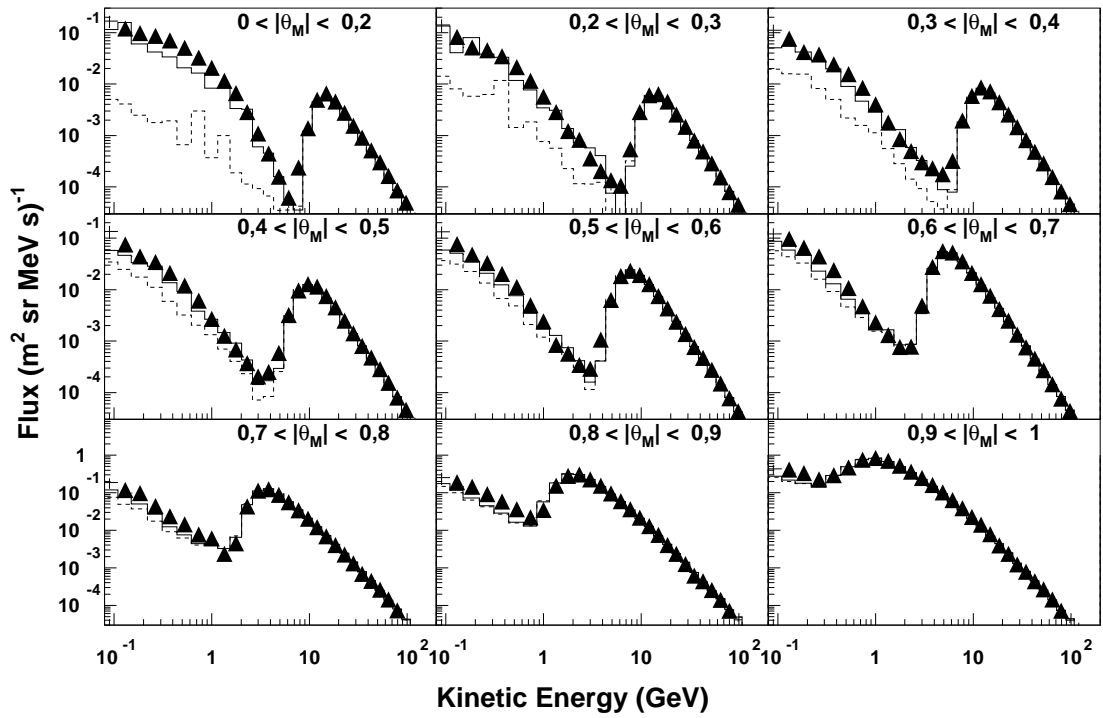}
 \caption{Downgoing proton flux, simulation(solid line) and the AMS data
(triangles); the dashed lines are described in the text. $\Theta_M$ is the
 geomagnetic latitude in radians.
}
\label{fig2}
\end{figure*}

\begin{figure}[!htb]
\center
\includegraphics[width=\textwidth]{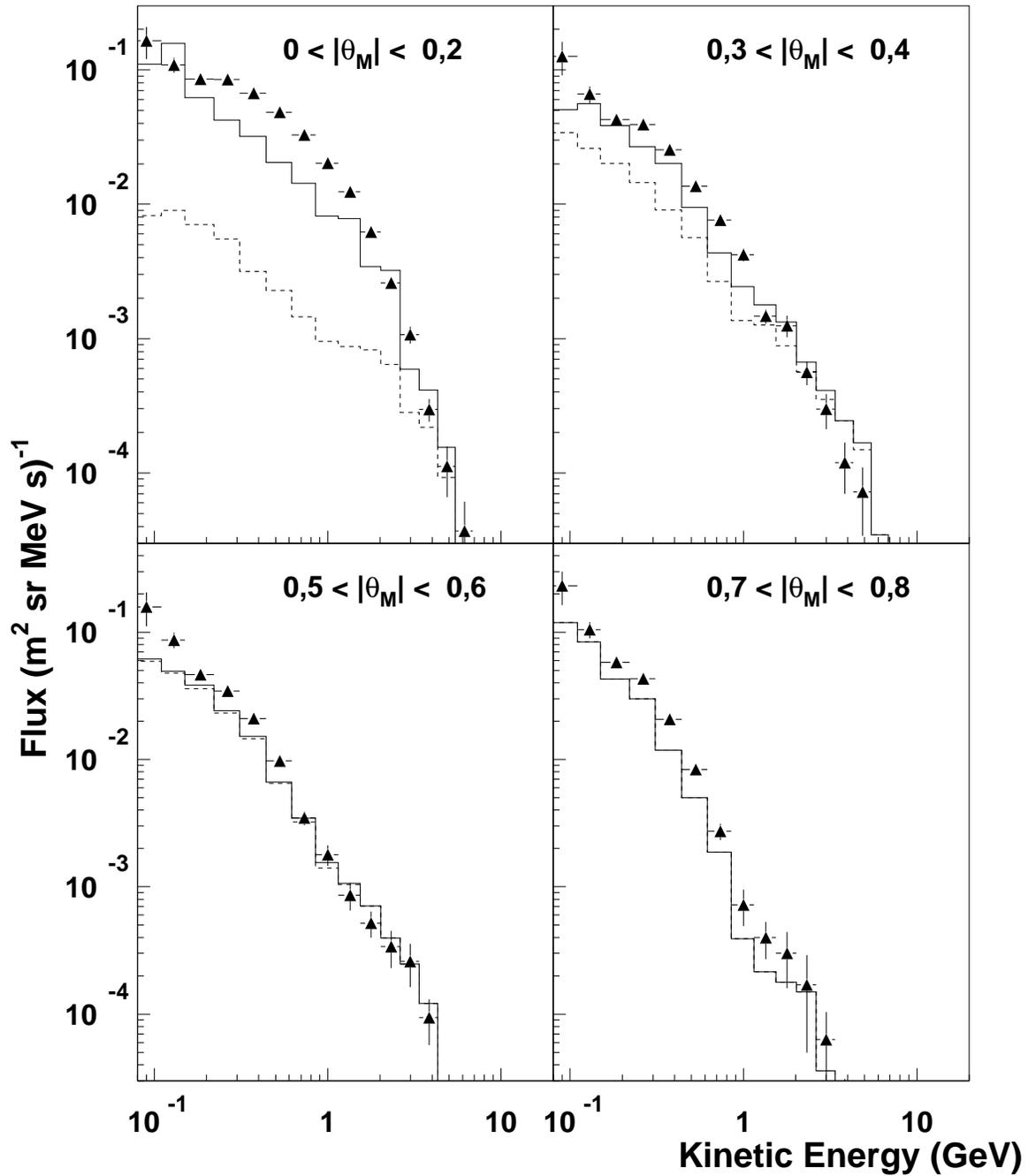}
\caption{Upgoing proton fluxes,  simulation~(solid line) and the AMS data
(triangles); the dashed lines are described in the text.}
\label{fig3}
\end{figure}

In Fig.\ref{fig2}, we show the comparison between the fluxes obtained with the
simulation and the measured AMS downgoing proton flux~\cite{AMSpra} in nine bins of
geomagnetic latitude ($\theta_{M}$) \cite{gustaffson}.
Fig.\ref{fig3} shows the same
comparison for the upgoing proton flux in four selected bins of $\theta_{M}$.

As seen in Fig.\ref{fig2}, the simulation well reproduces at all latitudes
the high energy part of the spectrum and the falloff in the primary 
spectrum due to the geomagnetic cutoff, thus validating the general 
approach used for the generation and detection, as well as 
the tracing technique.

A good agreement among data and simulation is also found in the under-cutoff 
part of the spectra. The small and systematic deficit which can 
be seen in the secondary component of the simulated fluxes is of the same
order of the expected contribution from the interaction of cosmic He and
heavier nuclei.

This flux is due to the secondaries produced in the atmosphere and that 
spiral along the geomagnetic field lines up to the detection altitude. 
Therefore it is sensitive to specific aspects of the interaction model 
and to the accuracy of the particle transport algorithm.

A correct quantitative prediction of this part of the spectra depends on 
the quality of low energy nucleon production both in terms of 
yield and energy distribution. This is in part due to the fragmentation 
of the target nucleus, and depends on the details of the nuclear physics 
algorithms describing excitation and break up. 

From the analysis of the motion of the secondary protons from their
production up to their detection, it can be pointed out that a 
fraction of the observed flux is due to a multiple counting of the
same particles.
Within the formalism of adiabatic invariants \cite{adiabatic}, it is 
seen that charged particles trapped in the geomagnetic field, i.e. the 
undercutoff protons, move along drift shells which can be associated
 with a characteristic residence time\footnote{The mean time after which a particle is absorbed into the atmosphere. In our case it represents the 
effective life time of the particle. }
that depends on the fraction of the shell located inside the Earth's atmosphere.
Thus, particles moving along {\it long-lived} shells have a
large probability to cross many times a geocentered spherical detector, 
while those moving along {\it short-lived} shells 
typically cross the detector only once. 

The drift shells crossing the AMS orbit, at an altitude of 400 km, 
are in general {\it short-lived}, however in the equatorial region
the {\it long-lived} shells are present as well \cite{fiandrini}.

In the following, we will indicate as the {\it real} proton flux
that one obtained by counting only once each particle crossing the detector:
its intensity is indicated by the dashed distributions in 
figs.~\ref{fig2}~and~\ref{fig3}. 

A quite relevant effect can be seen
in the equatorial region: there the AMS measurement indicates an 
important secondary proton flux while the {\it real} number of protons 
crossing the detector is more than one order of magnitude lower.
 At high geomagnetic latitudes, the solid and dashed lines tend to merge. 
The effect becomes negligible for $\theta_{M} > 0.8$. 

\begin{figure}[!htb]
\center
\includegraphics[width=\textwidth]{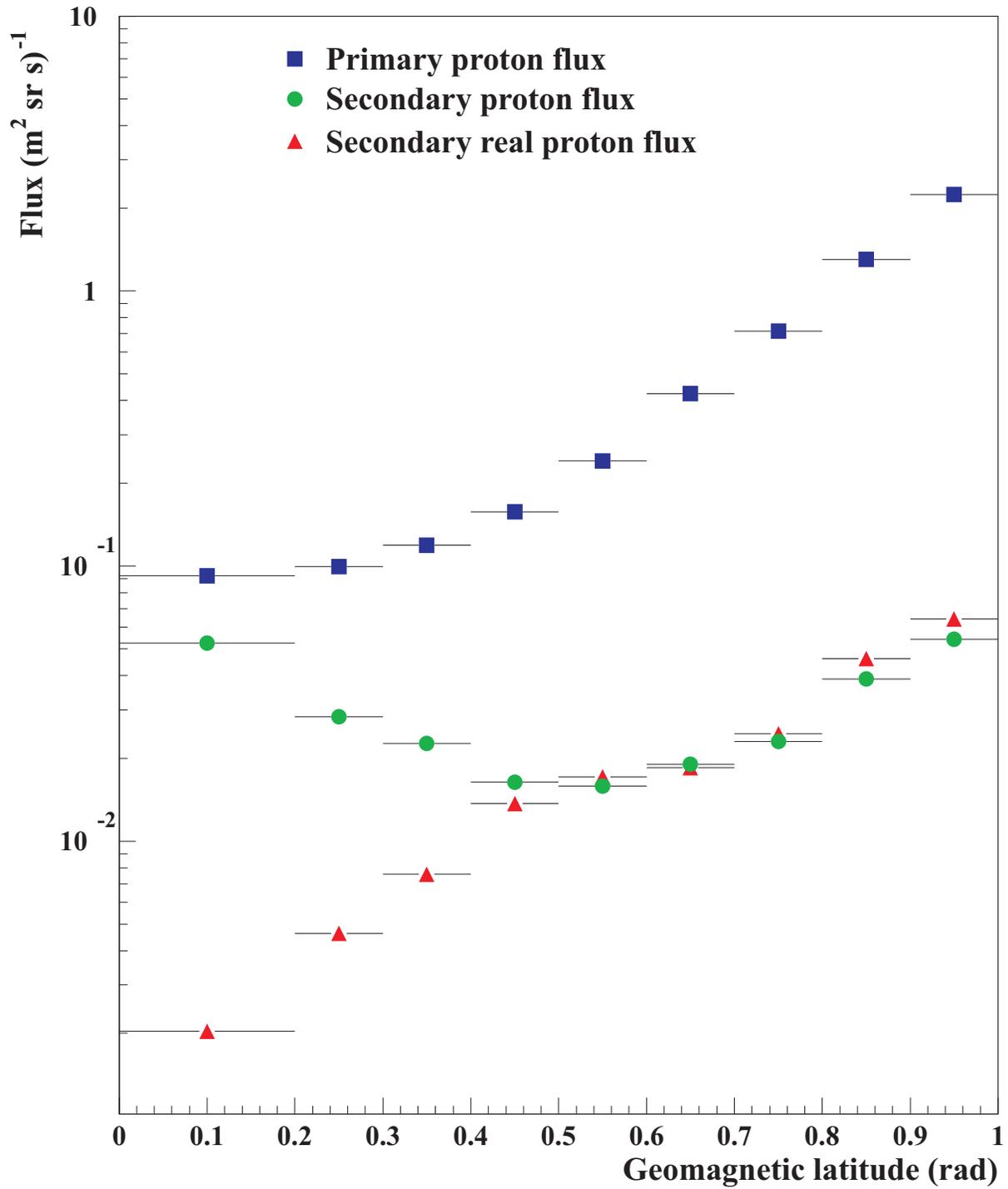}
\caption{ Proton fluxes in the AMS field of view as calculated with this 
simulation. The fluxes are integrated over the kinetic energy 
range $0.1-170~GeV$ and shown as a function of the geomagnetic latitude.}
\label{fig_fin}
\end{figure}

\begin{figure}[!htb]
\center
\includegraphics[width=\textwidth]{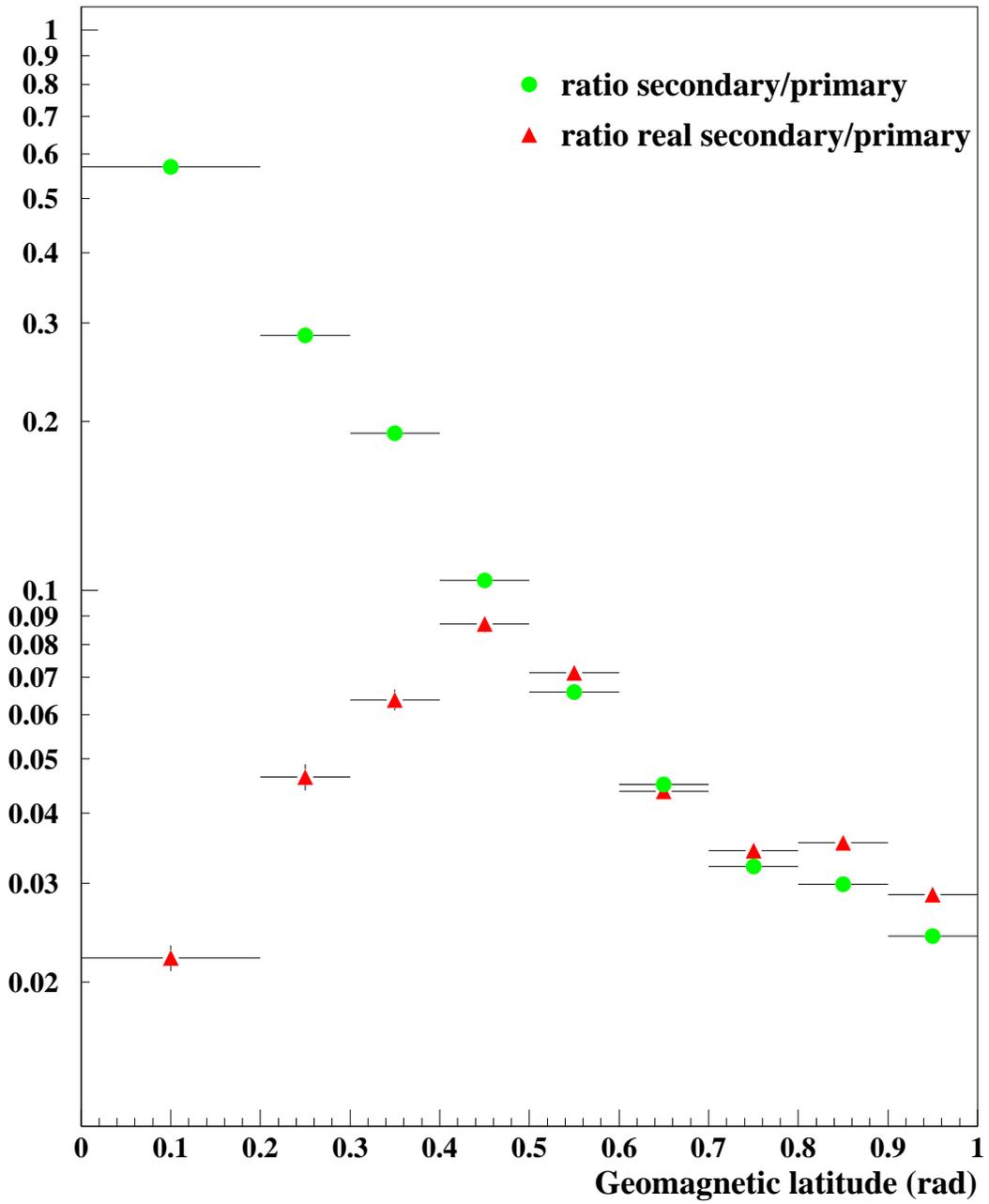}
\caption{ Ratios of the fluxes shown in Fig.\protect{\ref{fig_fin}}}
\label{fig_fin2}
\end{figure}

This can be better seen in Fig.\ref{fig_fin}, where the 
integral primary proton flux seen by AMS is shown as a function 
of geomagnetic latitude. The intensities of the {\it real} and  measured  
undercutoff fluxes are reported in the same plot for comparison and their 
ratio with the primary component  shown in Fig.\ref{fig_fin2}. A minor \
contribution from the undercutoff proton component can be therefore expected 
in the atmospheric shower development and neutrino production.

\begin{figure*}[!t]
\center
\includegraphics[width=\textwidth]{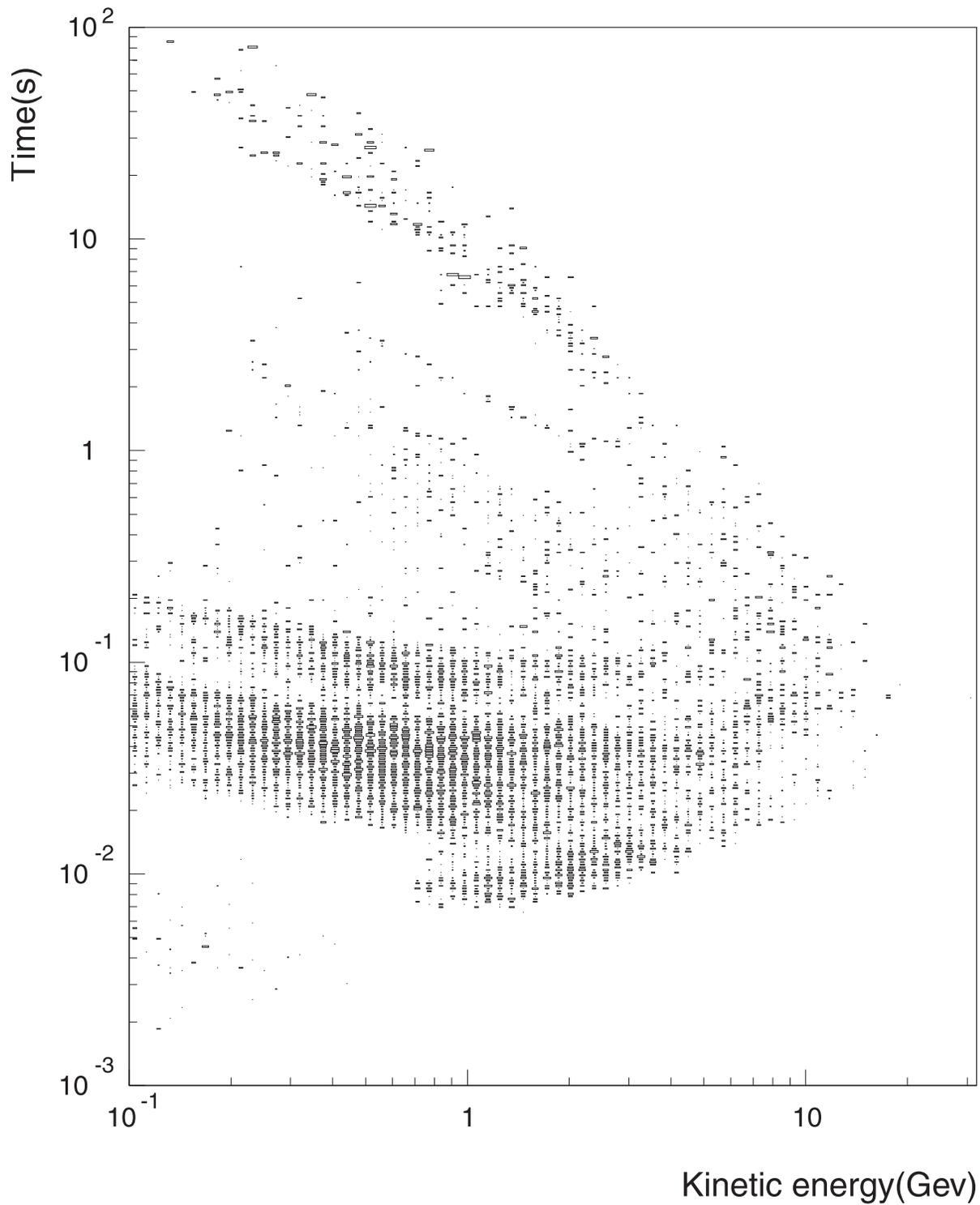}
\caption{ Life time versus kinetic energy for secondary protons produced in the
interactions of primary cosmic rays with the atmosphere. The protons are
detected at geomagnetic latitude $|\theta_{M}|< 0.4$ rad.}
\label{fig7}
\end{figure*}

In Fig.\ref{fig7}, the residence time is plotted versus the kinetic energy 
of the trapped secondary protons for $|\theta_{M}|\!<\!0.4$. In the scatter
plot it is possible to distinguish the populations corresponding
to {\it long-lived} and {\it short-lived} shells similar to those shown
in \cite{AMSele} for leptons.

\begin{figure}
\center
\includegraphics[width=\textwidth]{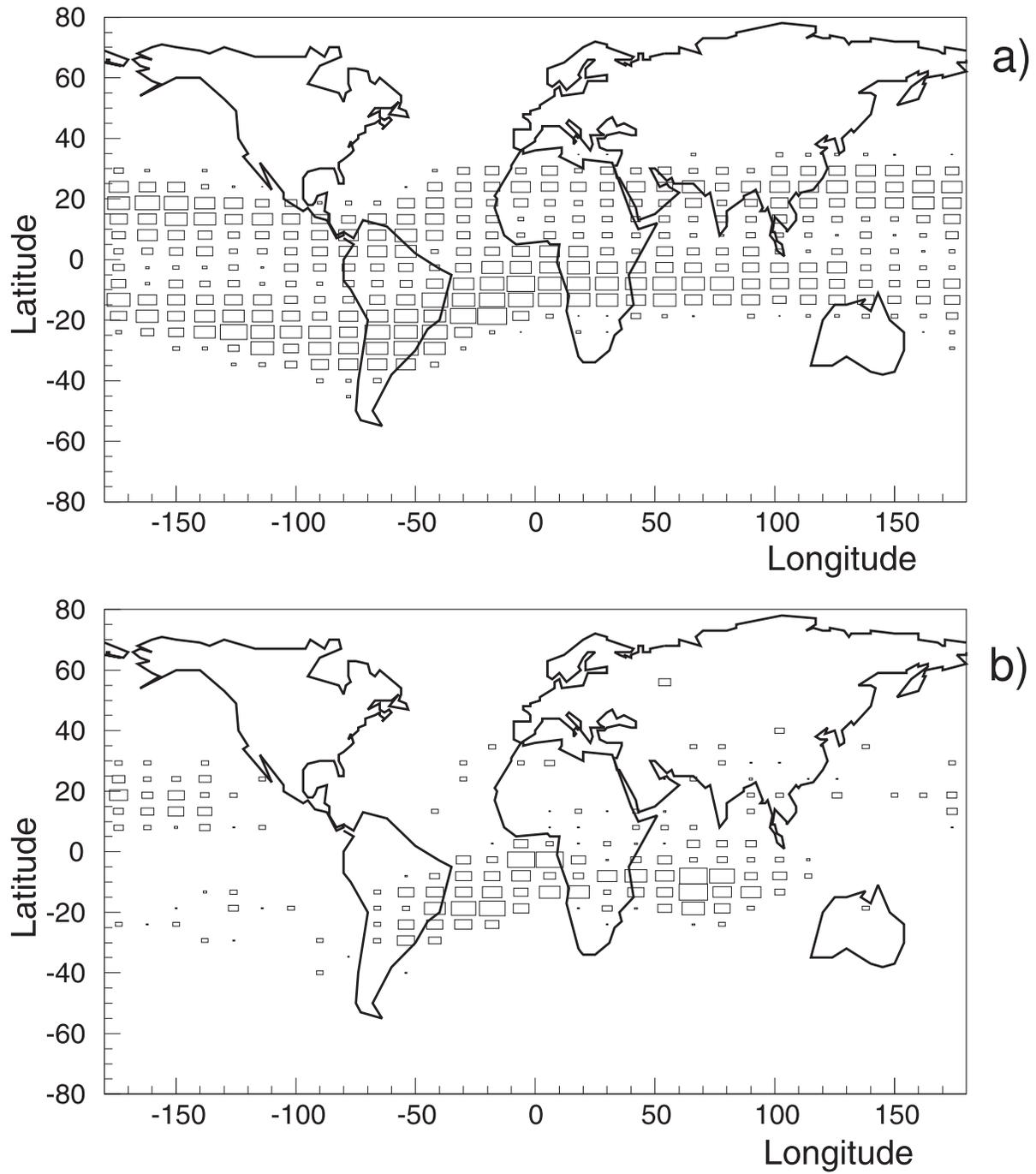}
\caption{Maps of secondary protons end points for  geomagnetic latitude
$|\theta_{M}|<0.4$ rad. Figure (a) live time $<0.3 s$ Figure (b) live time
$>0.3 s$}
\label{fig6}
\end{figure}

Fig.\ref{fig6} shows the distribution of trapped secondary proton end points for
$|\theta_{M}|\!<\!0.4$, Fig.\ref{fig6}a is for a lifetimes smaller than 0.3 s.,
while Fig.\ref{fig6}b is for a lifetimes greater than 0.3 s.. The 
end point distribution agrees with the location of the
intersections of the drift shells with the atmosphere
as experimentally
verified by \cite{AMSpra}, and   discussed in \cite{fiandrini}.

\subsection{Electrons and positrons}

\begin{figure}
\center
\includegraphics[width=0.7\textwidth]{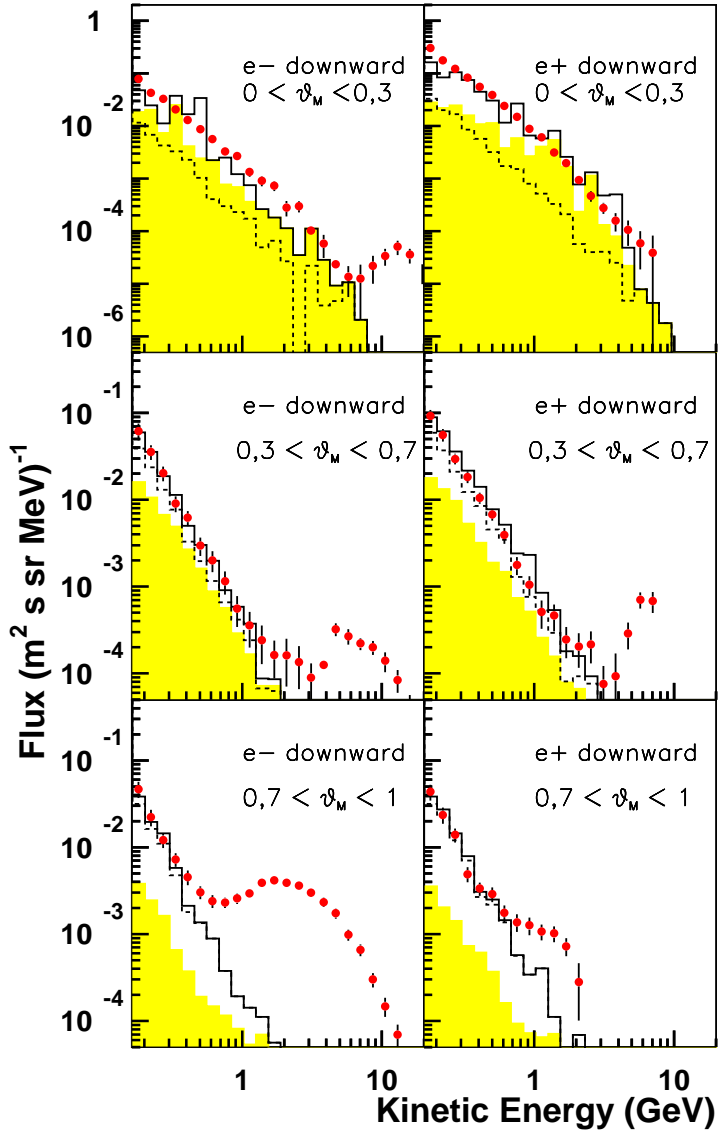}
\caption{Downgoing positron and electron fluxes in two regions of
geomagnetic latitude $\theta_{M}$, solid histogram (simulation)
black points (AMS data);
hatched histogram shows the positron and electron fluxes
produced by primary protons with $E_{k}>30~GeV$; the dashed line
distributions are described in the text.} 
\label{fig4}
\end{figure}

In Fig. \ref{fig4} we show a comparison 
of the simulated undercutoff electron and positron downgoing fluxes 
with the corresponding AMS measured fluxes~\cite{AMSele}.

We remind that the AMS positron measurement is restricted at
 energies below few GeVs, with a dependence of the maximum energy
 on the geomagnetic latitude which reflects the increasing 
proton background with $\theta_m$. 

A comparison of data and simulation in the high energy part of the
electron spectra is not possible, since the cosmic electrons have not
been used as an input in the current work. However, their contribution to
the cosmic rays reaching the atmosphere is $O(10^{-2})$  leading to a
negligible effect of in the generation of the undercutoff fluxes.

The simulation well reproduces the general behavior
of the undercutoff part of the spectra in terms of shape and intensity; 
a similar agreement is observed for the upgoing lepton spectra~(not shown).
The {\it real} lepton fluxes, corresponding to the {\it real} proton 
flux described earlier, are shown with the dashed line
 distribution in Fig. \ref{fig4}. As in the case of protons, a large effect
from multiple crossing is present going toward the equatorial region, 
more pronounced for the positron component.

As for the undercutoff protons, we would have expected a systematic deficit
in the simulated electron fluxes coming from the missing 
contribution of helium and heavier nuclei to the CR fluxes.
Subcutoff $e^{\pm}$ are mainly (97$\%$) coming from decays 
of pions produced in the proton collisions with the atmospheric nuclei:
charged pions contribute through the $\pi-\mu-e$ chain, while $\pi^{0}$
through $\pi^{0}\rightarrow \gamma~\gamma$
with subsequent e.m.  showers. 
The relative contribution of charged pions to the subcutoff 
electrons (positron) fluxes at AMS altitude in our simulation is found to be
$37\%$ ($47\%$), while the remaining $60\%$~($50\%$) appears to come from
$\pi^{0}$ production.  
This point deserves some considerations. The level of agreement in the
comparison of data and predictions for $e^{\pm}$ fluxes turns out to be an
important benchmark for 
the interaction model in view of a discussion on particle production in
atmosphere, since is strictly linked to the meson
production (mostly pions at this energy). This work
complements other studies oriented to the validation of
the FLUKA model in terms of particle yields. In particular,  
the quality of $\pi^{\pm}$ generation in our interaction model has been already
checked in \cite{batti01} through the  comparison with  muon fluxes
measurements  at different depths in atmosphere. 
The muons are from the charged pions 
decay chain and experimental data \cite{data_caprice} are well reproduced
by the simulation.
In the case of $e^{\pm}$ also $\pi^{0}$'s become relevant. 
Usually, when parametrized interaction models are used, like in the works of
ref. \cite{derome,Deromele}, the yield of $\pi^{0}$ is fixed assuming 
{\it a priori} an exact charge symmetry in pion production. In practice,
this is also
made necessary by the large errors that affect the scarce existing
experimental data on neutral pion production. Instead, in the case of a
microscopic interaction model like FLUKA,  there is no constraint of this
type, and the 
balance of $\pi^{0}$ vs. $\pi^{\pm}$ automatically emerges from the feature
of the model. Recently, it has been pointed out how in FLUKA there exists 
a significant violation of charge symmetry\cite{lipari_ven}: $\pi^{0}$'s
are in general more ($\sim$20\%) of the average of $\pi^{+}$ and $\pi^{-}$.
Technically this symmetry violation emerges in the hadronization of
color strings and this normally occurs also in other codes, like JETSET or
PITHYA\cite{Lund}. We cannot enter here in a detailed discussion of this
point, and we limit ourselves to say that there are reasons to believe
that, at least for laboratory energies below 100 GeV, the acceptable value
of charge asymmetry should be lower than that resulting from the present
version of the code. 
However, the comparisons of predictions to data discussed
in this work, combined with the mentioned work on atmospheric
muons\cite{batti01}, already tell that the predicted fraction of $\pi^0$ 
cannot be significantly wrong, 
although no definitive quantitative conclusion can be 
extracted, since the nuclear component 
has not been yet introduced in the primary spectrum.

\begin{figure}
\center
\includegraphics[width=\textwidth]{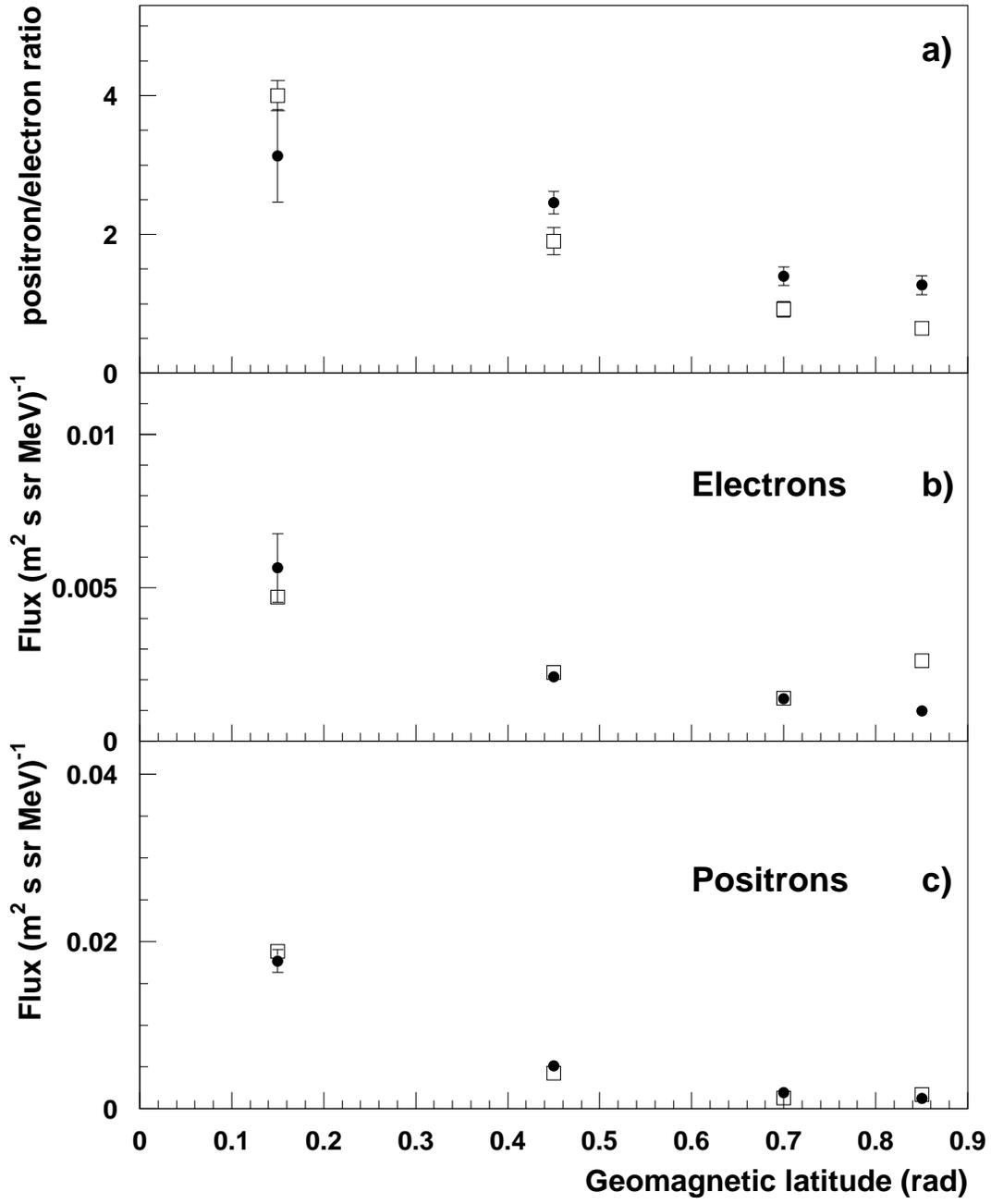}
\caption{ Electron (b) and positron (c) fluxes and their ratio (a)
integrated in the kinetic energy range $0.2-1.5~GeV$, as function of
geomagnetic latitude. Open squares (AMS data), black points (simulation).} 
\label{fig5}
\end{figure}

In Fig.\ref{fig5}b-c we show 
the integrated positron and electron downgoing fluxes
for the kinetic energy range $0.2\!-\!1.5~GeV$ as a function of $\theta_{M}$.
Their ratio is shown in Fig.\ref{fig5}a. One of the most remarkable
features of the AMS measurement is the large value of this ratio, when 
compared to the natural cosmic value, and its latitude dependence.
In Fig. \ref{fig4}, the contribution from primary protons with 
$E_{k}>30~GeV$ to the electron and positron fluxes is illustrated by the
filled area. We can notice that in the equatorial
region, the electrons are produced essentially by primary
protons with $E_{k}>30~GeV$ , while for the positrons lower energy protons
contribute as well. This distinction disappears at higher latitudes, where
positron and electrons are produced by the protons in the same energy range.

This behavior reflects the East-West asymmetry of the geomagnetic cutoff
on primary protons and the larger probabilities of escape from atmosphere
for secondary electrons(positrons) generated by Westward(Eastward) 
moving protons \cite{Lipari,Deromele}.

Positrons are preferentially injected on drift shells reaching the 
AMS altitude by eastward moving protons, which experience a lower rigidity
 cutoff than westward moving ones. This mechanism is more effective at the
equator, where the cutoff is larger and its asymmetry maximal, resulting 
in the excess of undercutoff positrons from low energy protons as indicated by
our simulation.
The cutoff mechanism becomes irrelevant at high latitudes, 
where any difference in positron and electron production should be
given instead by different $\pi^{+}/\pi^{-}$ production. \
Nor the data neither our
simulation indicate, within their uncertainties, a relevant
 charge asymmetry from this source.

\section{Conclusions}

The interactions of cosmic ray protons with the Earth's atmosphere 
have been investigated by means of a fully 3D Monte Carlo program.

The proton, electron and positron undercutoff flux intensities 
measured by AMS, as well as their energy spectra, have been correctly 
reproduced by our simulation

Geomagnetic effects, and in particular the east-west asymmetry 
in the cosmic protons rigidity cutoff, have 
been confirmed as the mechanism responsible for the measured 
excess of the positron component.

The main features of the geographical origin and residence time 
distributions for both protons and leptons have been replicated 
and the effect of multiple crossing of the detector by spiraling 
secondaries in the geomagnetic field briefly discussed. 

Our results indicate that the intensity of the undercutoff proton flux, 
when the multiple counting is taken into account, never exceeds a 
10\% of the cosmic proton flux, representing a negligible source
 for atmospheric production of secondaries. However, this aspect
will be object of further and more refined study in the future.

The analysis on the possible strategies to generate the cosmic
rays incoming flux  has shown the validity of a backtracing approach
as an accurate and highly efficient technique.  

This work provides also additional way to validate the features of
the adopted particle production model. In particular, 
the study of $e^{\pm}$ fluxes has revealed to be 
to be an interesting instrument to check the meson production in
primary interactions, and the results are satisfactory.
We believe that our
simulation, validated by the high statistic measurements of AMS, can 
be used to assess the radiation environment in near Earth orbit, and 
represents a valuable tool for a more accurate calculation of particle
fluxes in atmosphere.
\\

{\it This work has been partially supported by the Italian Space Agency (ASI)
under contract ARS-98/47.}


\newpage



\end{document}